\documentclass[11pt]{article}
\usepackage{graphicx}
\usepackage{fix-cm}

\begin{document}

\title{Exploring Network Economics}




\author{
{\rm Dah Ming Chiu, Wai Yin Ng} \\
{Department of Information Engineering} \\
{The Chinese University of Hong Kong}
}



\date{January 2011}

\maketitle

\begin{abstract}
In this paper, we explore what \emph{network economics} is all about, focusing on the interesting topics brought about by the Internet. Our intent is make this a brief survey, useful as an outline for a course on this topic, with an extended list of references. We try to make it as intuitive and readable as possible. We also deliberately try to be critical at times, and hope our interpretation of the topic will lead to interests for further discussions by those doing research in the same field.

\end{abstract}

\section{Introduction}
\label{intro}

\subsection{Motivation}

What is \emph{network economics}? It is a question worth pondering.

As a research topic, needless to say, it is an inter-disciplinary area between economics and networks. By \emph{networks}, we are referring mostly to the Internet, a network that permeates our society and daily life. The scope, however, can be much broader\footnote{Probably a more satisfactory perspecptive to economists.}, to include other forms of network such as transportation networks and social networks. We deliberately limit our scope so that we can focus our reflection on what network economic studies particularly have benefited the engineering and management of the Internet, and how the Internet may have changed the economic thinking or introduced new concepts. We also deliberately try to be somewhat critical as times, pointing out certain self indulgence or re-invention of the wheel.

The Internet is the biggest man-made object by many measures. Beginning as a computer network, it has rapidly become a global information infrastructure, a platform for exchange of information among people and machines all over the world. Commercialization in mid-1990s led not only to rapid expansion of the Internet as a computer network, it has led to radical change of what it is. \cite{FS2002} pointed out that what the new Internet connects are autonomous self-interested agents. Ownership, operation, and use by many self-interested, independent parties give the Internet the characteristics of an economy as well as those of a computer~\cite{FS2002}. Economic concepts of demand and supply, markets and mechanisms, competition and incentives, etc. have become new concepts for the Internet engineer in dealing with network traffic, resources, protocols, algorithms, and continual appearance of novel applications that bring along changes in traffic and user behaviour.

Our exploration of this topic is prompted by concrete motivations. At our department, we will start offering a graduate level course with the title \emph{Network Economics}. What would be suitable content for such a course, and how should we structure the syllabus to make it more interesting? Another question that came up recently is whether there should be a journal for this research topic, considering there is a steady stream of papers categorized to be \emph{network economics} without a publication venue of its own. While we do not try to explicitly answer these questions, we keep these questions in mind in our discussions.

A necessary disclaimer is that our exploration is unavoidably biased and non-comprehensive. Both the authors are engineers with only limited exposure to economics. Passing judgement over previous works on a research topic, though important in advancing the field, is subjective. While we read and edit\footnote{One of the authors, in his role as an editor for IEEE/ACM Transactions on Networking, helps process papers on network economics on regular basis.} many papers in this topic area, it is beyond our intention to do a comprehensive survey, but categorize and highlight only those studies that impress on us.

\subsection{Summary of the content of this paper}

Our approach is to focus on the network economic questions, starting from the simpler and more well formulated ones first. The methodologies, such as optimization, game theory and mechanism design, are introduced as they are called for.

We start with the study of network resource allocation policies. We discuss why the seemingly pure engineering issues become economic because of the need to use decentralized mechanisms and the need to consider fairness and efficiency. Examples of network mechanisms include congestion control and routing.

As the network becomes large, as is the case of the Internet, the way the network is connected and formed is not controlled by any single entity, but by autonomous entities that own and manage part of the network. We call them Internet Service Providers (ISPs), and the forces that lead the ISPs to peer and form the network are definitely to a great extent economic. Additionally, we examine other examples of network formation studies in Peer-to-peer networks and wireless networks.

Next we discuss \emph{network effect}. It is the benefit of a connected community in which any additional user generates benefit for all without cost. The Internet can be considered a platform with multiple types of stakeholders. Such an abstraction is useful for understanding how technology evolves and gets adopted, and the need for government regulation. For example, net neutrality, a hotly debated issue in the networking research community as it affects network architecture and design, is a complicated issue about maximizing network effect and how to balance its benefits among stakeholders.

Finally, we wrap up by briefly mentioning some other network economic topics that we have not included, making some observations about this field, and discussing promising future directions.

\section{Network Resource Allocation}
Someone wrote in Wikipedia: Economics is the social science that analyzes the production, distribution and consumption of goods and services. A large number of studies in the design, engineering, and management of networks (in particular the Internet) are considered from this perspective. 

\subsection{Policies and mechanisms of traffic control}
Perhaps the most salient example is the study of congestion control. On the surface, it can be treated as a pure engineering problem. It can be motivated by examining the detrimental effect congestion has on throughput, and the bandwidth waste and delay in repairing packet losses. Various solutions can be considered in terms of their ability to maintain throughput, and minimize retransmission and delay.

But before long, it was realized that economic issues are at the heart of the congestion control problem. A network provides service to many at the same time, and congestion is caused by excessive demand \emph{by many}. Who is responsible to alleviate it and receive reduced service? Chiu/Jain studied the notion of \emph{fairness} in network and computer system resource allocation~\cite{fairness}, and decentralized algorithms to do congestion control and achieve fair resource allocation~\cite{aimd}. These are examples of some early works on network resource allocation that exhibit economic thinking. Another important mechanism for congestion control is traffic routing, which will be discussed briefly later.
If we look into the economics literature, Varian wrote about fairness in the 70s \cite{hv_fair}, and studied the network congestion control problem via pricing in the early 90s \cite{pricing}. 

By late 90s, Kelly et al unified and generalized the previous work on network congestion control mathematically, as a decentralized mechanism to solve a utility maximization problem \cite{kelly}. This formulation allows the network feedback be naturally interpreted as either a congestion signal, or a shadow price; the notion of fairness be tied to the utility of the users/flows using the network \cite{mw98}; and the congestion control algorithm \cite{aimd} to be understood as an algorithm to solve an optimization problem. Amazingly, a network of users following the same (congestion control) mechanism, can be viewed as a smart and civilized community who maximizes their collective social welfare by maximizing their own utility reacting to given market prices\footnote{Note, users do not have to be price-takers, but try to out-smart others, which usually lead to less social welfare. This will be discussed in the next subsection.}, an aesthetically pleasing conclusion.

It is also interesting to point out the relationship of network resource allocation to the more general problem of combinatorial auctions (CA) \cite{ca}, studied in economics and theoretical computer science. The CA problem tries to \emph{allocate} different combination of bundles of goods to users via auction, and the users are modeled to have different utility for different bundles, making the problem quite complicated in general. In this sense, the network resource allocation problem is a special case of CA.

\subsection{Price of anarchy}
Since the congestion control mechanisms are decentralized, another valid and interesting \emph{economic} question is whether selfish behavior by users will render these mechanism ineffective? This question has been studied by economists and mathematicians before. There are two well-known works that vividly discuss this issue. In considering the sustainability of economic (or ecological) growth in the presence of selfish behavior, G Hardin wrote the famous essay on \emph{the Tradegy of the Commons} \cite{ttotc}. The \emph{Braess's Paradox} \cite{paradox} is even more to-the-point. By using a concrete example, the paradox illustrates that selfish behavior make users not able to benefit from additional network resources.

To characterize the spread between the suboptimal resource allocation due to selfish behavior and the optimal social welfare, Papadimitriou coined a term - \emph{price of anarchy} \cite{papadimitriou}. For this problem, Roughgarden et al derived some fundamental results for routing \cite{roughgarden}, providing the complete solution for Braess' paradox; Johari et al studied the congestion control game and derived the price of anarchy in that case \cite{johari}.

It is worth pointing out that the study of selfish behavior, though based on a game-theoretic analysis of an operating equilibrium, treats the users as a homogeneous population. Hence, the specific motives of users are ignored, and the scenario is simplistic compared to the game-theoretic analysis of stakeholders in networks in the next section. Nonetheless, the price of anarchy results are useful as a general guide to the design of network mechanisms, to the extent the model of the selfish motives and behaviors are plausible.

\subsection{QoS and network pricing}
There is a huge literature on the study of network \emph{Quality of Service} (QoS). QoS is also about resource allocation policies in networks, about favoring, or providing guarantees to certain classes traffic. Since these policies and mechanisms are rarely deployed in the Internet, it would be speculative to discuss the network economics issues thereof\footnote{Rather, the lack of deployment makes a good economic question for study.}.

Network pricing is a related topic amassing research effort more recently. For the sake of economic equilibrium, traditional price theory mandates pricing that increases with quantity demanded of the product in question. That would mean usage-based pricing in network services, even congestion-based pricing. In the presence of significant positive network effects, however, perhaps a more important consideration is the size of the network, as that directly determines the value of the network to users. So there is incentive for service providers to charge a flat-rate price. This was realized by AT\&T long time ago for telephone networks, see \cite{history} for an excellent account of the history of pricing in communications systems. This is indeed what is practiced in most parts of the Internet, and other telecommunication services. We will return to the importance of network size (also known as network effect) for network economics in a later section.

Furthermore, we want to point out an important distinction between two notions of network pricing, namely pricing for cost recovery, and pricing for profit. Broadly speaking, resource pricing for cost recovery is an integral part of network design, and therefore engineering, the breakeven charge plan necessary to settle the cost account of an engineering design for its implementation and operation. Technical assumptions about network traffic and utilization are made based on design requirements. In contrast, product pricing for profit is business concern beyond breakeven, for which business assumptions such as market size and demand elasticities are made based on market research instead. While traffic models and market models may look similar in analytical forms, they are categorically distinct and should be constructed and validated where they belong. We opine that, without proper market research, theoretical work tinkling with product pricing for profit is a questionable feat.


\subsection{Paris metro pricing}
There is, however, a neat idea with a nice tie to some economic theories worth a description here.  The scheme is known as \emph{Paris Metro Pricing}, proposed by Andrew Odlyzko, for use in a congested network to provide a differentiated service \cite{paris}. The idea came from a practical system, as the name suggests, Paris Metro\footnote{Presumably at one point in time. It is not clear if Paris Metro still implements this form of service, but it is still in operation in other parts of the world, for example in Hong Kong's train system.}. It works like this. Each car of the metro is either designated as First Class or Second Class. If you enter the first class, you pay an additional charge to the normal fare (for Second Class). The beauty is that the mechanism self-adapts with users' reaction to congestion and the additional charge. When the congestion level of First Class reached a sufficiently high level (though still better off than that of the Second Class), some people would start moving to the Second Class cars for a better deal, and vice versa. For reasonable assumptions of continuity of users' tolerance to congestion and additional charges, it is intuitive to see that the system converges to an equilibrium state.

The key economic question is whether Paris Metro Pricing achieves higher social welfare, and profits for the operator of the metro (or Internet). Apparently, there were contradicting answers to this question in the literature for several years. In turns out that indeed you can reach both conclusions depending on how you model users' reaction to congestion. The situation is analogous to two roommates sharing a small apartment which can be either arranged as a studio (a single large room) or partitioned into two smaller rooms. You ask different people whether they prefer the former or the latter, and you will get different answers. One type of people is referred to as \emph{partition-preferring} whereas the other type is referred to as \emph{multiplexing-preferring}. In a recent paper \cite{pmp}, we used this classification of users to develop sufficient conditions for Paris Metro Pricing to be viable, i.e. able to gain more social welfare, and/or provider profit. The paper contains additional results to unify previous works on Paris Metro Pricing.

This study of Paris Metro Pricing also makes ties to the classic work by Hotelling \cite{hotelling} when he studied product differentiation (primarily among competitors, which can also be applied to a single provider). Hotelling models a range of users with different preferences, by using the example of a beach with two ice-cream stands and a distribution of how beach-goers are spread out on the beach. He is able to derive the popularity of each ice-cream stand for different locations of placing them, and reach the conclusion that two competing ice-cream stands would put themselves as close to each other as possible.  This Hotelling model of diverse users can also be applied to derive the social welfare, and evaluate the best resource allocation for a single provider system. Indeed, it is useful for a variety of network economic analysis. This is the reason we make a special point of introducing it here.

\section{Economics of network formation}
While the last section concerns resources allocation in a network, the question considered here is how (large) networks are formed and maintained in the first place.

\subsection{Internetworking - transit and peering}
By internetworking, we mean the formation of a network of networks. The Internet is a network formed when a hierarchy of backbone and access providers interconnect with each other. Any such network provider faces two basic questions: (1) whether connection is desirable; (2) if connection is desirable, which other providers he should connect with. The first question seems a no-brainer for a profit-seeking network provider serving customers who desire connectivity, but a real question nevertheless. Conceivably, if his customers desire connectivity among themselves only, connection with the outside world is not necessary. Otherwise, it spells the fundamental condition of an Internet network provider. (Even when customers desire Internet connection but obtain it elsewhere by multihoming, the provider may do without interconnection. However, such a provider would, by definition, not be an Internet network provider.)

The second basic question is more involved, with both engineering and economics considerations. Transit and peering are two different categories of interconnection. Except for backbone providers in the topmost tier, a network provider needs to procure transit service from another provider, often one in its immediate upper tier. He may also peer with other providers, often in the same tier so as to save transit cost and improve service quality at the same time. However, providers are often competitors also and handing customer traffic over to one another is not without hesitation. Service level agreement (SLA) for peering is problematic. Settlement is a more tangible sub-problem here: traffic demand and requirements in the two directions of a peering connection differ in general and how the imbalance should be accounted for properly is still an unsettled issue. As a result, two-way peering tends to be restricted to providers of similar scale, with simple or no settlement arrangement for traffic imbalance. In comparison, multi-way peering is common in practice with an Internet Exchange (IX) as an intermediary, operated often by a public or impartial organization. Such arrangement, connecting network providers of a large city for instance, offers benefits to all participants significant enough to ignore any imbalance.


Although in real life ISP peering is based on complicated business considerations, game theory can help us gain insights into some fundamental questions. For example, \cite{ss2006} used evolutionary game theory to illustrate that it is in the long term interest of the ISPs to be connected. In \cite{m2010}, authors argued that ISPs should cooperate and showed how they can share the benefit of a connected and optimally operated network fairly based on the concept Shapley value in cooperative game theory.

\subsection{Peer-to-peer networking}
In hindsight, the appearance of Napster in 1999 was an epochal event. Since then, peer-to-peer (p2p) file sharing has proliferated and rapidly become a major end-user application, and a bandwidth hog that poses serious network management challenges. Such peer-to-peer networking leads to two major areas of concern.

First, there are networking issues for p2p end-users (clients), namely, how the network should be formed for a specific instance or, for an individual client, which other clients it should connect with. Sustenance of a p2p network, for file sharing for instance, relies on client contribution of processing and bandwidth by uploading data for one another. The key problem is free-riding behaviour \cite{AH2000}, when a client downloads without uploading. Incentive mechanism needs to be in place to balance clients' download and upload shares. The phenomenal success of Bittorent hinges on the use of simple bilateral tit-for-tat rule \cite{C2003}, by which a client would choke its upload to another client who is found not uploading in return. The clients form a market through which demand and supply match up for exchange of file data. Different p2p protocols are implementations of different distributed exchange mechanisms, for which bilateral barter is the simplest.

Second, network providers serving p2p clients are faced with a host of engineering and economic challenges due to two major reasons. Firstly, p2p traffic is often locality-unaware overlays, being routed oblivious of the underlay physical networks \cite{RFI2002}. Being locality-unaware means being cost-unaware in the choice of routes, resulting in inefficient overuse of expensive transit bandwidth for instance. Secondly, content distribution by p2p clients shifts traffic away from content providers who are often charged by traffic volume, to end-users who are often charged flat rate, which translates to revenue loss to network providers \cite{KRP2005}. In principle, the two-way tussle between network provider and p2p clients may be mitigated with better coordination, when some sort of locality signalling \cite{AFS2007} facilitates the alignment of p2p traffic overlays with the network underlay. However, The three-way interaction among content providers, end-users and network providers can be very complicated with competing economic interests.

\subsection{Ad hoc wireless networks}
In recent years, there has been a huge level of interest in the study of wireless networks. Besides the more practical use of wireless networks as access networks (e.g. cellular networks or Wifi networks), the other paradigm is known as \emph{ad hoc networks}, a multi-hop wireless network spanning potentially a large area formed by autonomous nodes. These nodes discover each other, and figure out how to connect to each other to form a network. Each node may serve as a relay for other nodes some times. This process can be very complicated since wireless transmissions within a close vicinity interfere each each other. Besides such technical challenges, the nodes, as autonomous decision makers, also worry about whether they should free-ride, or try to optimize for the social welfare. The shared wireless spectrum becoems an economy, in which each node evaluates its strategy in the game it plays with all the other nodes. This combination of technical and economic issues provide a rich set of scenarios for research studies. But these ad hoc networks are hardly adopted in practice, and there is no clear business model to validate the model with. The economic analyses of such ad hoc wireless networks, in our opinion, become rather meaningless.

\section{Network effects}
Network effects have been studied as an economic phenomenon for decades, especially since deregulation of US telecommunications industry in 1980's. Antitrust investigation in the software industry, such as that of Microsoft before Internet times, put network effects into public limelight. Like every popularized concept, the term has been overused and misused, to the extent that even plain old economy-of-scale that economizes unit cost of production is sometimes mistaken as a network effect which scales value-to-customer as the consumer base increases. Liebowitz further emphasizes that the term network externalities should be reserved to mean network effects which benefits are not captured by any player or market and remain ``external¡¨ \cite{LM1994}. For instance, directory service of a centralized telephone network would help user discover more useful numbers and connect more. The network effect, not realized otherwise, is said to be internalized by the directory service provider in this case, and shared with all users. Without the directory service, the network effect would have remained an externality. The distinction between internalized network effects and non-internalized network externality is indeed important as it carries policy implications. Network externalities are often indication of threat or opportunities beyond the stakeholders' control, such as when directory service is poorly provisioned and users are not connecting as much as they would be otherwise, it may be a case of ``market failure" that warrants intervention, for instance, when public funding is sought for provisioning a directory service.

Congestion, discussed in a previous section, which may cause traffic volatility, even instability, is a network effect of the negative kind. An important strategy of congestion control is to prevent the spread of congestion, by dropping packets at the congested link to signal link users to back off. This containment strategy, restricting the negative effects to those responsible, has an economic interpretation. The network effect is said to be {\it internalized}. Otherwise, when congestion spreads, the network effect becomes network externality, causing impact to innocent flows elsewhere.

While negative network effects such as congestion poses threat to the Internet, positive network effects are prime drivers of its rapid growth. There are two basic principles, both economical. First, network components are used in combination; for instance, any path between two network users is a tandem of links. Addition of a component, such as a link, increases the number of combinations available, by creating new alternative paths. Second, the value of the Internet to any user increases with addition of every new user, which means a larger community to connect with, and potentially more content being shared. In either case, the side benefits to others are welcome network externalities. Such externalities in aggregate are powerful incentive to all sorts of businesses and innovations, which in turn generate more Internet traffic, and congestion. In broad term, the tussle between positive and negative network effects is a fundamental limit to Internet's growth.

\subsection{How much network effect}
It is noteworthy that different ways to quantify network effect have been a topic of debate. The most well-known is \emph{Metcalf's Law}, which has become a folklore. The law says that the value of a communications network is proportional to the square of the number of users connected to it. Basically, this is characterizing the positive network effect of a communications network as the maximum number of other users a user can connect to.

Several years after the bust of the dotcom euphoria, Andrew Odlyzko et al wrote an article to refute Metcalf's law \cite{odlyzko}. By blaming the financial down-turn on Metcalf's law, this paper drew quite a bit of attention. The basic idea is that not all other users generate the same amount of positive network effect. In the end, it is more appropriate to give $log(n)$ credit to the network effect experienced by each user, thus the total network effect would be $n log(n)$. They reached the $log(n)$ amount in several ways. You can say each user is only likely to know $log(n)$ other users; or you can even rank the values of other users by \emph{Zipf's law} to arrive at the $log(n)$ conclusion. Their article is quite an interesting read.


\subsection{Economic theory of network effects}
It was Katz and Shapiro's timely paper \cite{katzshapiro} in 1985, prompted by the then seachange in telecommunications industry and consumerization of computing, that initiated the economic theory of network effects. They studied the nature of competition in many technology settings: they tried to explain when multiple technical solutions compete, why sometimes the best technology may not win (for example Beta versus VHS, and the operating systems battles); and when do competitors agree to standardize their technologies \cite{katzshapiro}. They succeeded in explaining many such contemporary phenomena in terms of network effect. In particular, they show that multiple equilibria are possible, among which joint adoption of product standard is one. The extent of network effect, much of which predicated on consumersÕ expectation, is a key determinant. As such, it explained, perhaps even predicted, fierce competition and innovation in influencing consumersÕ expectation and early behaviour, exemplified by generous support for software developers, giveaway trial versions, the freemium approach, etc.

\begin{figure}
\begin{center}
  \includegraphics[width=0.7\textwidth]{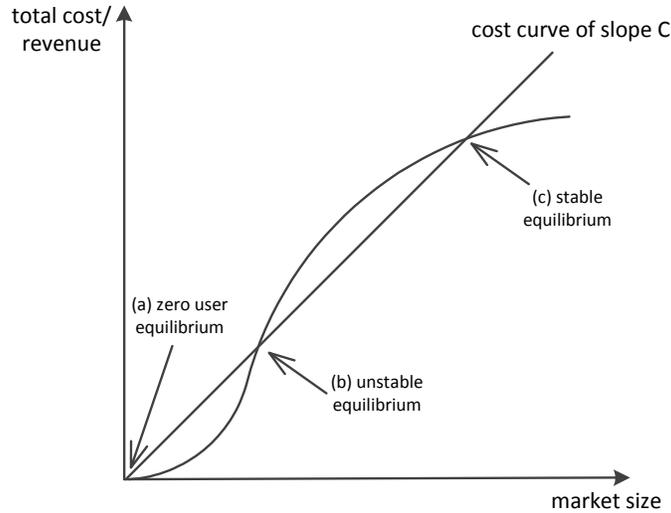}
\end{center}

\caption{Market equilibria with network effects} 
\label{neteffect}       
\end{figure}

One of the most familiar result concerns the viability of a service (or product) in view of network effect. Suppose there is a unit cost of $c$ in delivering the service to each user. The total cost for $n$ users is represented as a straight line. Due to network effect, the value of the service (product) is represented as a sigmoidal curve whose value is initially below the cost line, but quickly rises above the cost line. Eventually, as $n$ approaches total market size, it naturally tapers off. See Fig.~\ref{neteffect} for an illustration. In a market with free competition, revenue (user value) equals cost. At least three equilibria exists: (a) the zero user solution; (b) the unstable solution when the total network effect first covers all the costs for $n$ users; and (c) when users and network effect both increase until the total network effect sustains a much larger user base. They are the three places the value curve crosses the cost line. The equilibrium (b) is not stable because any change in the user number will cause the system to converge to one of the other two equilibria. There is a description of this in \cite{economides}; the same observation comes up in many other studies involving network effect.

Later, Katz and Shaprio also studied models on \emph{systems competition}, that is, when a service (or system) is offered by two or more products, such as hardware and software, how network effect affect products across products. For example, a software solution for a particular hardware platform can benefit from the increasing user base of that hardward platform, and vice versa \cite{katzshapiro2}.

\subsection{Net neutrality}
Perhaps the most controversial topic in Internet economics is the still on-going debate surrounding net neutrality. From a business model point of view, Internet Service Providers (ISPs) are keen in competing for business opportunities that can benefit from the network effect, such as content distribution, on-line social networks or games - by, if not directly entering these businesses, by extracting profits from these services. The Federal Communications Commission (FCC) of the USA, however, wants to keep the Internet as open as possible as a platform for innovation and free competition. To ensure this, FCC wants the ISPs to stay completely neutral in offering their service to all users.

It is not surprising to see economists at opposite ends when arguing about government policies. This is less common with (networking) engineers who usually converge to the same models and same conclusions. But network effect (although it has the word \emph{network} in it) is more about economics than networks. Many economists argue that excessive regulation, albeit in the good name of net neutrality, can induce market inefficiencies \cite{netneutrality} while others assert that when net neutrality is compromised, with excessive service differentiation and price discrimination profitable to some but constricting to the community's growth, network effect will be gravely undermined to the detriment of all eventually.

\begin{figure}
  \begin{center}
  \includegraphics[width=0.9\textwidth]{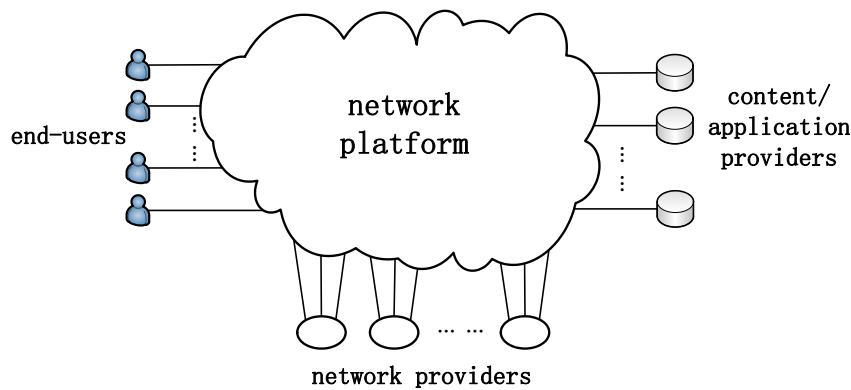}
  \end{center}

\caption{Internet as a two-sided platform} 
\label{twoside}       
\end{figure}

The concept of a multi-sided market, which captures the notion of platform in digital economy, has provided useful elaboration of network effects in this connection. The Internet is an exemplary multi-sided market \cite{RT2003}, with the Internet a platform around which different types of users gather, including network providers, content/application providers, and end-users. Besides the two basic kinds of network effects, a multi-sided market exhibit also {\it cross-side} network effects. Consider content/application providers and end-users. The benefit to a content/application provider, in business terms, increases with the number of end-users. Also, the benefit to an end-user, in access to content and services, also increases with the number of content/application providers. Economides formulated a simple two-sided market model of the Internet with explicit cross-side network effects in both directions, as illustrated in Fig.~\ref{twoside}, then went on to show that net neutrality can ensure the right mix of such network effects so that social benefit is maximized \cite{netneutrality2}.

\section{Concluding remarks}

\subsection{Some topics not included}
There are many other topics we could have included, but for a variety of reasons were left out. The most likely reason is due to the limitations of our background. We both had a primarily engineering training, with some reading in economic topics. We hoped to point out more useful economic concepts and re-inventing-of-the-wheels related to them, than we managed to do.

There are certain topics, while quite interesting to us, seem more relevant to specific businesses and their operations than an economic issue, for example various e-commerce topics, or Ad auctions. They certainly can be included as part of a network economics course, and will likely receive a lot of interest.

Other than \emph{net neutrality}, we have not discussed much on government policies. There are certainly a lot of material in this area, including the management of frequency spectrum, regulation of content distribution, and others. Some of these issues become as much related to national security as economics.

Finally, economists have also studied many interesting problems arisen due to various kinds of information technologies, not necessarily networks. For example, the issue of how to price informational goods, and how it is different from physical goods? In the section on \emph{network effects}, we touch upon some of these more economic (rather than network) issues. Some of these topics are arguably quite related to the Internet and can be included in a course on network economics as well.

\subsection{Observations}
The engineering and management of the Internet and its value-added services inevitably involve economic considerations. The abstraction, generalization and analysis of these considerations crystalize into network economics (or more specifically, Internet economics). Some of the topics are induced by the Internet, and are unique to the networks; whereas some others are special applications of generic economic principles.

The study of network economics helps us to understand the demand for network connectivity, bandwidths and services, so that we can make better design of the Internet. For example, Kelly used utility functions to model users' demand for network bandwidth; Metcalf, Odlyzko and Economides used network effects to help characterize the consumer economy-of-scale on network services. 

The interesting problems of network economics - we hope we picked some to discuss in this paper - are usually inspired by the real tussles between economic stakeholders, even if the problems have gone through some abstraction. The moral is, we need to talk to real-life service operators, and study real business models and popular phenomena, to understand what and where the interesting problems are.

As networks, applications and network services continue to expand, so will the study of network economics.  We think network economics makes an interesting inter-disciplinary course for student majoring in computer network, Internet technologies and related areas. Such a course can be offered at either graduate level, or undergraduate level. The book and course by Easley and Kleinberg \cite{ek} is a good example for an undergraduate level course of this type\footnote{This course is not only focused on network economics, but also on network science applied to some other new Internet services and phenomena.}.

We believe there is value for a journal devoted to network economics. This will help keep research papers in this topic area easier to find, and build a small research commitee to foster collaboration. The NetEcon and WINE workshops, which have been run for several years, are already quite helpful. To be successful, however, it must be truly inter-disciplinary, drawing researchers from both economics as well as networking/engineering.

\subsection{Promising future directions}
Moving forward, we see that Internet will connect not only our computers, but also our phones, TVs and many other devices (e.g sensors). The dispute (and cooperation) between network delivery providers, content providers and other value-added service providers will continue; so the debate on net neutrality has only just begun. We expect to see interesting developments here.

Various efforts (in US, Europe as well as in Asian countries) are devoted to future Internet architectures. The viability of these new designs must make economic sense. For this reason, we anticipate interesting network economic studies in relation to the future Internet efforts. Two thoughts came up in our brainstorming. First, when we worked on service differentiation (QoS) more than 10 years ago, the dotcom bubble led to over-provisioning, which rendered QoS unnecessary. But network QoS may become necessary again, due to the increase in traffic caused by new services (e.g. high-definition video) and the access network turning mostly wireless which has finite bandwidth to share. Second, as the whole world is becoming more energy and environment conscious nowadays, this may become an important consideration, which is somewhat economic.

Finally, as we write, it seems to us that this whole research area is waiting for a more overarching theory to emerge. Perhaps a Nobel prize will be awarded for network economics some day?

\bibliographystyle{IEEEtran}
\bibliography{netecon}   

%
%


\end{document}